\documentclass[prc,superscriptaddress,unsortedaddress,twocolumn,showpacs]{revtex4}
\usepackage{amsmath}
\usepackage{amssymb}
\usepackage{times}
\usepackage{mathrsfs}
\usepackage{multirow}
\usepackage{bm}
\usepackage{wrapft}
\usepackage{color}
\usepackage{graphicx}
\usepackage{ulem}

\def\fun#1#2{\lower3.6pt\vbox{\baselineskip0pt\lineskip.9pt
\ialign{$\mathsurround=0pt#1\hfil##\hfil$\crcr#2\crcr\sim\crcr}}}

\newcommand{\beq}{\begin{equation}}
\newcommand{\eeq}{\end{equation}}
\newcommand{\bea}{\begin{eqnarray}}
\newcommand{\eea}{\end{eqnarray}}

\newcommand{\bfi}[1]{\mbox{\boldmath $#1$}}
\newcommand{\bfis}[1]{\mbox{\boldmath ${\scriptstyle #1}$}}

\newcommand{\vK}{{\bfi K}}

\newcommand{\vR}{{\bfi R}}

\newcommand{\vka}{{\bfi \kappa}}

\newcommand{\viK}{{\bfis K}}

\begin{document}

\title{Probing three-nucleon-force effects via ($p$,$2p$) reactions}

\author{Kosho Minomo}
\email[]{minomo@rcnp.osaka-u.ac.jp}
\affiliation{Research Center for Nuclear Physics, Osaka University, Ibaraki 567-0047, Japan}

\author{Michio Kohno}
%\email[]{kohno@rcnp.osaka-u.ac.jp}
\affiliation{Research Center for Nuclear Physics, Osaka University, Ibaraki 567-0047, Japan}

\author{Kazuki Yoshida}
%\email[]{yoshidak@rcnp.osaka-u.ac.jp}
\affiliation{Research Center for Nuclear Physics, Osaka University, Ibaraki 567-0047, Japan}

\author{Kazuyuki Ogata}
%\email[]{kazuyuki@rcnp.osaka-u.ac.jp}
\affiliation{Research Center for Nuclear Physics, Osaka University, Ibaraki 567-0047, Japan}

\date{\today}

\begin{abstract}
We propose to use proton knockout reactions ($p$,$2p$) from a deeply bound orbit
as a new probe into three-nucleon-force (3NF) effects.
The remarkable advantage of using ($p$,$2p$) reaction is that we can choose an appropriate
kinematical condition to probe the 3NF effects.
We analyze ($p$,$2p$) reactions on a $^{40}$Ca target within the framework of distorted-wave impulse approximation with
a $g$-matrix interaction based on chiral two- and three-nucleon forces.
The chiral 3NF effects significantly change the peak height of
the triple differential cross section of ($p$,$2p$) reaction.
We also clarify the correspondence between the ($p$,$2p$) cross sections and
the in-medium $pp$ cross sections.
\end{abstract}

\pacs{21.30.Fe, 24.10.Eq, 25.40.-h}
%21.30.Fe  Forces in hadronic systems and effective interactions
%24.10.Eq  Coupled-channel and distorted-wave models
%25.40.-h  Nucleon-induced reactions

\maketitle

%{\it Introduction.}
Three-nucleon forces (3NFs) are one of the essential elements
which dictate a variety of the dynamics in few-nucleon systems, finite nuclei, and nuclear matter.
Without 3NFs, even the saturation property of nuclei can not be explained,
and equation-of-state of neutron star is unrealistic~\cite{akm98}.
For nuclear structure calculations,
the 3NF gives the last piece to reproduce binding energies of light nuclei~\cite{wir02}.
For nucleon-deuteron elastic scattering,
3NF effects are clearly observed in cross sections at middle and large angles,
in which the two-nucleon force (2NF) is relatively small~\cite{wit98}, and are also detected in spin observables~\cite{sek11}.
In heavier systems, 3NF effects may be represented by
the density-dependence of the nucleon-nucleon (NN) effective interaction (the $g$-matrix interaction)~\cite{fur08,raf13}.
Very recently, nucleus-nucleus elastic scattering is investigated as a probe of 3NF effects at high density~\cite{fur16}.
3NF effects reduce the cross sections at large scattering angles significantly,
where higher than normal density is produced in the overlapping area between the projectile and target densities.

Besides three-nucleon correlations via ordinary NN interaction,
3NFs may arise from intermediate excitations of non-nucleonic degrees of freedom, typically an isobar $\Delta$.
A pioneering work for two-pion exchange 3NFs by Fujita and Miyazawa~\cite{fuj57}
was followed by various phenomenological models, such as
Tucson-Melbourne~\cite{coo79} and Urbana~\cite{pud97} models, which have been used frequently to
investigate the roles of 3NF effects in nuclear systems.
In the last two decades, the theoretical description of 3NF has progressed considerably
by chiral effective field theory (Ch-EFT)~\cite{epe09,mac11,mac16}, in place of phenomenological 3NFs.
Ch-EFT enables us to organize two-, three-, and many-body forces consistently and systematically.
The chiral interactions are applied to microscopic calculations of
few-nucleon scattering~\cite{wit16}, finite nuclei~\cite{nog06,ots10}, and nuclear matter~\cite{heb10,sam12,koh13}.
Furthermore, the $g$-matrix interaction including the chiral 3NF contributions has been constructed
and applied to describe many-body reactions~\cite{min14,toy15,min16}.

In this Letter, we investigate chiral 3NF effects in nuclear many-body reactions,
in particular, the proton knockout reaction ($p$,$2p$).
This reaction is suitable for studying effective nucleon-nucleon interactions
since ($p$,$2p$) reactions at intermediate and high incident energies can be
regarded as two-proton quasi-elastic scattering in the nuclear medium.
In fact, ($p$,$2p$) reactions from deep-hole states were used to investigate
the density-dependence of nucleon-nucleon effective interactions~\cite{nor00,hil06}.
We analyze ($p$,$2p$) reactions within
the framework of distorted-wave impulse approximation (DWIA)~\cite{cha77,cha83}
with the $g$-matrix interaction based on the chiral 2NF and 3NF.
DWIA is well established as a standard method
to describe knockout reactions in which the energy- and momentum-transfer are large.
A noticeable advantage of using knockout reaction is that
we can choose an appropriate kinematical condition which
satisfies the momentum condition for quasi-NN scattering
by changing the experimental condition.
This situation is quite distinct from nuclear structure and nuclear matter calculations.
We elucidate the roles of 3NF effects through the density dependence of the $g$-matrix interaction and
demonstrate the possibility of probing 3NF effects via knockout reactions.

%{\it Framework}
Let us consider the reaction in which the proton (0) with
the incident momentum $\vK_0$ (in the unit of $\hbar$)
hits the internal proton in the single-particle state denoted by $\alpha$ in $^{40}$Ca target,
and, two protons (1 and 2) are emitted with the outgoing momenta $\vK_1$ and $\vK_2$.
In DWIA in the asymptotic momentum approximation~\cite{yos16},
the triple differential cross section (TDX) of the ($p$,$2p$) reaction is given by
\begin{widetext}
\bea
\frac{d^3\sigma}{dE_1^{\rm L} d\Omega_1^{\rm L} d\Omega_2^{\rm L}}
&=&C^2S~F_{\rm kin}C_0
\bigg|\int d\vR~\chi_{\viK_1}^{(-)}(\vR)\chi_{\viK_2}^{(-)}(\vR)
\langle\vka'|g(k_{\rm F}(R),E)|\vka\rangle
\chi_{\viK_0}^{(+)}(\vR)\phi_{\alpha}^{}(\vR)\bigg|^2.
\label{tdx}
\eea
\end{widetext} 
Here, $E_1^{\rm L}$ is the outgoing proton energy and
$\Omega_1^{\rm L}$ and $\Omega_2^{\rm L}$ are the solid angles of the outgoing two protons in the laboratory frame.
$C^2S$ is a spectroscopic factor corresponding to a single-particle state of the proton in $^{40}$Ca,
and $F_{\rm kin}$ and $C_0$ are the kinematical factors defined in Ref.~\cite{oga15}.
$g(k_{\rm F}(R),E)$ is an effective nucleon-nucleon interaction, that is, a $g$-matrix interaction in the nuclear medium;
it depends on the incident energy $E$ of the bombarding proton and the local Fermi-momentum $k_{\rm F}(R)$.
$\vka$ ($\vka'$) is a relative momentum between the two interacting protons in the initial (final) state.
$\chi_{\viK_i}^{(\pm)}$ ($i=0, 1,$ and $2$) is a distorted wave of particle $i$,
and $\phi_{\alpha}^{}$ is a single-particle wave function of the knocked-out proton in the initial state.

The $g$-matrix is evaluated in nuclear matter with a Fermi momentum $k_{\rm F}$.
The details of obtaining the $g$-matrix based on the chiral interactions~\cite{epe05} is given in Ref.~\cite{koh13}.
For applying the $g$-matrix to finite nuclei, we adopt the so-called local density approximation.
The local Fermi-momentum $k_{\rm F}(R)$ is estimated with the matter density profile of $^{40}$Ca.
We use the proton density deduced by electron scattering~\cite{vri87} including
the finite-size effect of the proton charge by the standard unfolding procedure~\cite{sin78},
and it is assumed that the neutron density is the same as the proton one.

In this study, we adopt the chiral interactions of the J$\ddot{\rm u}$lich group~\cite{epe05},
with the cutoff parameters $\Lambda=450, 550$, and $600$ MeV,
and the low-energy constants for the NN sector are $c_1=-0.81$~GeV$^{-1}$, $c_3=-3.4$~GeV$^{-1}$, and $c_4=-3.4$~GeV$^{-1}$.
The other low-energy constants, $c_D$ and $c_E$, are determined to reproduce reasonably
the empirical saturation point in symmetric nuclear matter within the lowest-order Bruckner theory;
the resultant values of the saturation density $\rho_0^{}$, the binding energy per nucleon $E/A$ at $\rho_0^{}$,
and the incompressibility $K$ are tabulated in Table~\ref{tab1}.
We discuss theoretical uncertainties coming from this cutoff dependence of chiral interactions later.
%----------------------
% Table 1
%----------------------
\begin{table}[bth]
\caption{
Low-energy constants of the N$^2$LO chiral 3NF contact terms
and the corresponding values of the saturation density $\rho_0^{}$, the energy per nucleon $E/A$ at $\rho_0^{}$,
and the incompressibility $K$ for each cutoff parameter $\Lambda$.
}
\begin{center}
\begin{tabular}{c|ccccc} \hline\hline
 $\Lambda$ (MeV) & $c_D$ & $c_E$ & $\rho_0$ (fm$^{-3}$) & $E/A$ (MeV) & $K$ (MeV) \\ \hline
 $450$ & $-2.5$ & $0$    & $0.170$ & $-15.4$ & $193$ \\
 $550$ & $-2.5$ & $0.25$ & $0.152$ & $-12.6$ & $147$ \\
 $600$ & $-2.5$ & $0.5$  & $0.148$ & $-11.4$ & $126$ \\ \hline\hline
\end{tabular}
\label{tab1}
\end{center}
\end{table}
%----------------------

In the asymptotic momentum approximation~\cite{yos16},
assuming the conservation of the total momentum of NN system,
the kinematics of NN system, $\vka$ and $\vka'$, used in Eq.~(\ref{tdx}) are determined by
the asymptotic kinematics in the initial and final states.
The off-the-energy-shell (off-shell) properties are thus fully respected within the DWIA framework.
For a ($p$,$2p$) reaction with large binding energy comparable to the incident energy,
which is the case in Fig.~\ref{fig2} below,
the off-shell effects are non-negligible as discussed in Ref.~\cite{red70}.
The relativistic correction for the kinematics of the NN system is included
by the so-called M{\o}llor factor in this calculation.

Although 3NF gives influences in several ways in describing the knockout
process, we focus on its effects in the $g$-matrix interaction.
The distorted waves and single-particle wave function
are thus calculated with phenomenological local potentials.
We employ global proton optical potentials parametrized by Koning and Delaroche~\cite{kon03}.
The spin-orbit part is omitted for calculating distorted waves, because
it is not essential for describing unpolarized observables.
Effects of nonlocality are taken into account in both single-particle and scattering wave functions
by means of the Perey factor with the range of nonlocality $\beta=0.85$ fm~\cite{per62}.
We calculate the single-particle wave function $\phi_{\alpha}^{}$
with the mean-field potential obtained from the analysis of electron elastic scattering~\cite{elt67}.

\begin{figure}[tbp]
\begin{center}
\includegraphics[width=0.45\textwidth,clip]{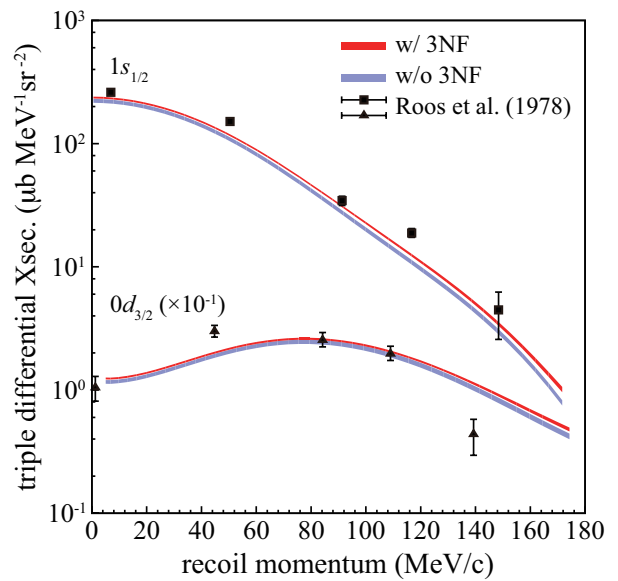}
\caption{(Color online)
Triple differential cross sections of $^{40}$Ca($p$,$2p$)$^{39}$K at the incident proton energy of $148.2$ MeV,
as a function of the recoil momentum of the residue.
The dark- and light-shaded bands correspond to the results with and without 3NF, respectively.
The width of each band represents the uncertainties coming from the cutoff dependence 
of chiral interactions.
The experimental data are taken from Ref.~\cite{roo78}.
The cross sections for $0d_{3/2}$ are multiplied by $10^{-1}$ for visibility.
}
\label{fig1}
\end{center}
\end{figure}

%{\it Results and discussions.}
In Fig.~\ref{fig1}, we show the TDX of
$^{40}$Ca($p$,$2p$)$^{39}$K at the incident proton energy of $148.2$ MeV,
as a function of the recoil momentum of the residue.
The knocked-out proton occupies $0d_{3/2}$ or $1s_{1/2}$ in the initial state
and the emitted angles of the two protons are fixed
at $\theta_1^{\rm L}=44^\circ$ and $\theta_2^{\rm L}=39^\circ$.
The experimental data are taken from Ref.~\cite{roo78}.
The dark- and light-shaded bands correspond to the results with and without 3NF effects, respectively,
and the width of each band represents the theoretical uncertainties coming from
the cutoff dependence of the chiral interactions.
The theoretical results include spectroscopic factors
$2.58$ for $0d_{3/2}$ and $1.03$ for $1s_{1/2}$,
which are deduced from the $(e,e'p)$ reaction analysis~\cite{kra89};
these values have been confirmed to be consistent with those deduced from
the ($d$,$^3$He) reaction analysis~\cite{kra01}.
%----------------------
% Figure 1
%----------------------
%----------------------

The difference between the results with and without 3NF is small
and the cutoff ambiguities are also negligibly small both for $0d_{3/2}$ and for $1s_{1/2}$.
Since the protons bound in those outer-most orbits are knocked out
mainly in the surface region of $^{40}$Ca,
the reaction process is hardly affected by the 3NF effects, as discussed later in detail.

The theoretical results well reproduce the experimental data in both cases of $0d_{3/2}$ and $1s_{1/2}$.
In Ref.~\cite{roo78}, the estimated spectroscopic factors were $4.0$ for $0d_{3/2}$ and $1.4$ for $1s_{1/2}$.
The main difference between their and our results comes from the distorting potentials.
Because the modern optical potentials adopted~\cite{kon03} are more well-tuned,
the present framework is expected to be more reliable.

In Fig.~\ref{fig2} we predict the TDX of
$^{40}$Ca($p$,$2p$)$^{39}$K at the incident proton energy of $150$ MeV,
as a function of the recoil momentum of the residue.
The $0p_{3/2}$ proton is supposed to be knocked out.
The emitted angles are fixed as $\theta_1^{\rm L}=30^\circ$
and $\theta_2^{\rm L}=40^\circ$.
$E_1^{\rm L}$ is changed from $18$ MeV to $73$ MeV.
The spectroscopic factor is assigned to be $4$, which corresponds to the single-particle limit for $0p_{3/2}$ orbit.
%----------------------
% Figure 2
%----------------------
\begin{figure}[tbp]
\begin{center}
\includegraphics[width=0.45\textwidth,clip]{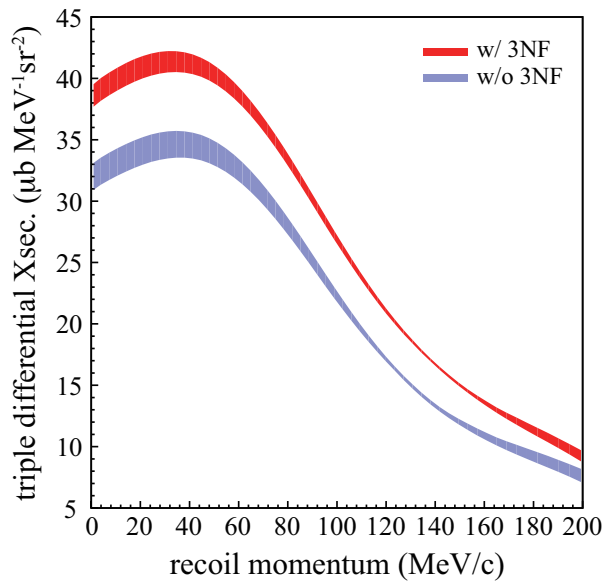}
\caption{(Color online)
Triple differential cross sections of $^{40}$Ca($p$,$2p$)$^{39}$K at $150$ MeV
as a function of the recoil momentum of the residue.
The $0p_{3/2}$ proton is knocked out and the emitted angles are fixed as $\theta_1^{\rm L}=30^\circ$
and $\theta_2^{\rm L}=40^\circ$.
The meaning of the shaded bands is the same as in Fig.~\ref{fig1}.
}
\label{fig2}
\end{center}
\end{figure}
%----------------------

In this case, we observe large enhancement of the cross sections by the 3NF effects.
At the peak, the cross section increases by $20$\%, while
the theoretical uncertainty coming from the cutoff dependence of chiral interaction is as small as $4$\%.
The increase in the cross sections is mainly due to the enhanced tensor component by the chiral 3NF effects.
The kinematics of the NN system corresponding to the peak in Fig.~\ref{fig2} is estimated as
$\kappa'=0.983$~fm$^{-1}$, $\kappa=1.46$~fm$^{-1}$, and $\theta_{\kappa\kappa'}=78.7^\circ$.
Here, $\theta_{\kappa\kappa'}$ is the angle between $\vka$ and $\vka'$, that is, the scattering angle of the two protons.
It is well known that noncentral components are dominant at around $90^\circ$ for unpolarized $pp$ cross sections.
Thus the 3NF effects can be clearly detected in the knockout reaction from a deeply bound state.
It should be noted that, for elastic scattering, the 3NF effects are generally discussed at large scattering angles,
in which cross sections are small.
The situation for the ($p$,$2p$) reaction is very different from this.

When a head-on collision between two protons occurs,
the relative momenta $\kappa$ and $\kappa'$ becomes large;
The non-zero recoil momentum region in Figs.~\ref{fig1} and \ref{fig2} corresponds to this condition.
Even in that case, we confirmed that $\kappa$ and $\kappa'$ are smaller than
the lower limit of the cutoff scale in the present calculations.

\begin{figure}[tbp]
\begin{center}
\includegraphics[width=0.45\textwidth,clip]{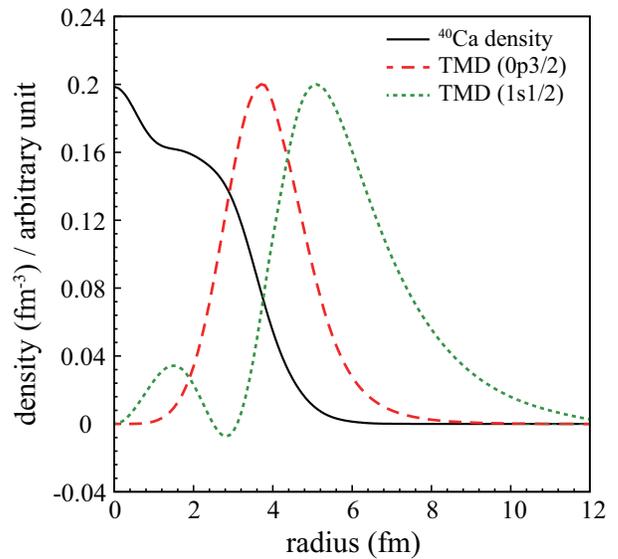}
\caption{(Color online)
The matter density profile of $^{40}$Ca (solid line) and the TMD corresponding to
the knockout reactions from $0p_{3/2}$- (dashed line) and $1s_{1/2}$-orbit protons (dotted line).
The TMDs are shown in arbitrary units.
The mean densities calculated with dashed and dotted lines are
$\bar{\rho}~(1s_{1/2})=0.022~{\rm fm}^{-3}$ and $\bar{\rho}~(0p_{3/2})=0.076~{\rm fm}^{-3}$, respectively.
}
\label{fig3}
\end{center}
\end{figure}

To evaluate the in-medium effect for the reactions discussed,
we introduce the transition matrix density (TMD) $\delta(R)$ defined by
\bea
\int dR~\delta(R) \propto \frac{d^3\sigma}{dE_1^L d\Omega_1^L d\Omega_2^L},
\eea
which indicates a transition strength as a function of $R$~\cite{hat97}.
In Fig.~\ref{fig3}, we show the matter density profile of $^{40}$Ca and
the $\delta(R)$ corresponding to the knockout reaction from $1s_{1/2}$- and $0p_{3/2}$-orbit protons.
The $\delta(R)$ is shown in arbitrary units.
%----------------------
% Figure 3
%----------------------
%----------------------
One sees that the TMD for the $1s_{1/2}$ knockout (dashed line) is more peripheral than that for $0p_{3/2}$ (dotted line).
Calculating an expectation value of the matter density with the TMDs,
we can estimate a mean density $\bar{\rho}$ for each kinematics,
which stands for a typical density assigned for
the density dependence of the $g$-matrix interactions.
The evaluated values of $\bar{\rho}$ near the quasi-free condition, that is,
at the zero recoil momentum in Fig.~\ref{fig1}, and the peak in Fig.~\ref{fig2}
are $\bar{\rho}~(1s_{1/2})=0.022$~${\rm fm}^{-3}$ and $\bar{\rho}~(0p_{3/2})=0.076$~${\rm fm}^{-3}$.
The $0p_{3/2}$ knockout gives the information on 3NF effects at around half normal-density,
while only the low density region is relevant in the $1s_{1/2}$ knockout.

\begin{figure}[tbp]
\begin{center}
\includegraphics[width=0.45\textwidth,clip]{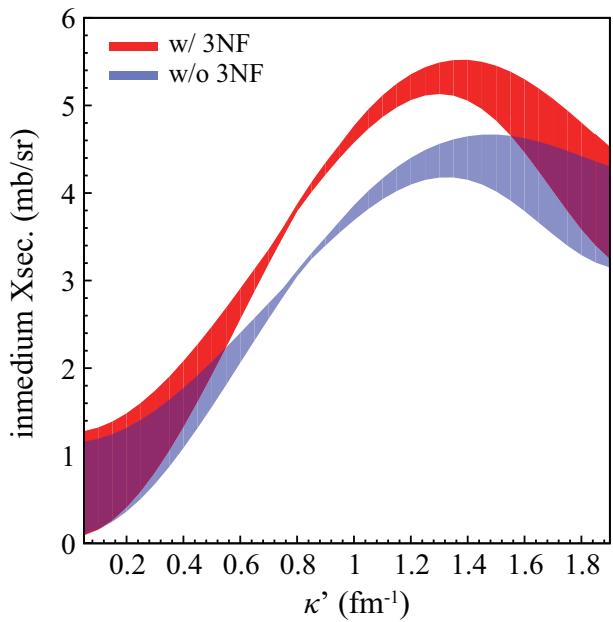}
\caption{(Color online)
Unpolarized in-medium $pp$ cross sections as a function
of the relative momentum in the final state $\kappa'$,
at $E=150$~MeV, $k_{\rm F}^{}=1.1$~fm$^{-1}$, $\kappa=1.5$~fm$^{-1}$, and $\theta_{\kappa\kappa'}=80^\circ$.
The meaning of the shaded bands is the same as in Fig.~\ref{fig1}.
}
\label{fig4}
\end{center}
\end{figure}

To discuss the correspondence between the 3NF effects and the reaction observables more closely,
we focus on the in-medium $pp$ cross section itself.
Figure~\ref{fig4} shows the unpolarized in-medium differential cross sections of $pp$ scattering,
which is proportional to $|\langle\vka'|g(k_{\rm F},E)|\vka\rangle|^2$, as a function of $\kappa'$.
We fixed the kinematics as $E=150$~MeV, $k_{\rm F}^{}=1.1$~fm$^{-1}$, $\kappa=1.5$~fm$^{-1}$, and $\theta_{\kappa\kappa'}=80^\circ$.
These set together with $\kappa'=1.0$~fm$^{-1}$ correspond to the kinematics of the NN system at around the quasi-free condition in Fig.~\ref{fig2}.
We expect that this behavior of the NN elementary process is directly reflected
in the TDX in Fig.~\ref{fig2}, and in fact, the $24$\% increase and $4$\% uncertainty of
the $pp$ cross section at $\kappa'=1.0$~fm$^{-1}$ are quite consistent with the amount of the change in TDX.
It indicates clearly that the ($p$,$2p$) reaction is suitable for probing the 3NF effects.
%----------------------
% Figure 4
%----------------------
%----------------------

It is noted that the small cutoff dependence of the chiral interactions
in Fig.~\ref{fig2} is rather accidental.
As seen in Fig.~\ref{fig4}, the 3NF effects and cutoff uncertainty depend on the off-shellness.
Conversely, this might allow us to investigate the off-shell properties of nucleon-nucleon interaction
in the DWIA framework.

%{\it Summary.}
In summary, we have analyzed ($p$,$2p$) reactions within the DWIA framework
with the chiral $g$-matrix interaction including 3NF and off-shell effects.
The remarkable advantage of using ($p$,$2p$) reaction is that the kinematics of
the NN elementary process is under control.
We showed the present framework is highly predictable
and found the 3NF effects are clearly observed by choosing an appropriate kinematical condition.
As a result, we can propose to use ($p$,$2p$) reactions from a deeply bound orbit
as a new probe of 3NF effects.

To clarify the chiral 3NF effects by comparing the present results with the experimental data,
the spectroscopic factors should be determined.
However, the uncertainties of analysis of deeply-bound nucleon knockout reactions are expected to be somewhat large
even for ($e$,$e'p$) reactions~\cite{mou76}.
If it is difficult to discuss the absolute values of cross sections,
we can consider spin observables such as vector and tensor analyzing powers of knockout reactions.
These observables are defined by the ratio of the cross sections with different spin-directions so that
we can avoid the uncertainty of the absolute values.
It is also interesting to compare ($p$,$2p$) and ($p$,$pn$) cross sections 
for examining the isospin dependence of 3NFs.
In future, it could be possible to put limits on the uncertainty of 3NFs and the off-shell properties
from many-body reaction observables.

%{\it Acknowledgements.}
The authors thank T. Wakasa, T. Noro, K. Sekiguchi, and M. Toyokawa for fruitful discussions.
The numerical calculations in this work were performed at RCNP.
This work is supported in part by
The Grant-in-Aid for Scientific Research
(Nos. JP15J01392, JP25400255, JP16K05352, JP16K05353, and JP16K17698)
from Japan Society for the Promotion of Science (JSPS).

%%--------------------------------------------------------------------%%
%%                           References                               %%
%%--------------------------------------------------------------------%%

\end{document}